\documentclass[11pt]{article}

 \font\gotb eufm10 scaled \magstep1
\newcommand{\bb}{\bibitem}
\newcommand{\cc}{\cite}
\newcommand{\vp}{\varphi}
\newcommand{\sss}{\sigma}
\newcommand{\al}{\alpha}
\newcommand{\pP}{P(\aA)}
\newcommand{\lt}{\left}
\newcommand{\rt}{\right}
\newcommand{\lll}{\lambda}
\newcommand{\F}{\hat F}
\newcommand{\D}{\hat D}
\newcommand{\R}{\hat R}
\newcommand{\s}{\hat S}
\newcommand{\K}{\hat K}
\newcommand{\I}{\hat I}
\newcommand{\Q}{\hat Q}
\newcommand{\Aa}{\bar A}
\newcommand{\A}{\hat A}
\newcommand{\B}{\hat B}
\newcommand{\HH}{\hat H}
\newcommand{\p}{\hat p}
\newcommand{\po}{\Psi_0}
\newcommand{\aA}{\tilde A}
\newcommand{\tvp}{\tilde \vp}
\newcommand{\AAA}{\hbox{\gotb A}}
\newcommand{\QQ}{\hbox{\gotb Q}}
\newcommand {\QQQ}{\hbox {\gotb Q}_{\xi}}
\newcommand {\vx}{\vp_{\xi}}
\newcommand {\vpx}{\vx(\A)}
\newcommand{\BB}{\hbox{\gotb B}}
\newcommand{\bea}{\begin{eqnarray} \label}
\newcommand{\eeq}{\end{equation}}
\newcommand{\beq}{\begin{equation} \label}
\newcommand{\eea}{\end{eqnarray}}
\newcommand{\nn}{\\ \nonumber}
\newcommand{\rr}[1]{(\ref{#1})}

 \author{D.A.Slavnov}

\title{ Quantum mechanics as a complete physical theory }

   \date{}

\begin{document}

  \maketitle
\begin{center}
{\it  Department of Physics, Moscow State University,\\
 Moscow 119899, Russia. E-mail: slavnov@goa.bog.msu.ru }
\end{center}
 \begin{abstract}

We show that the principles of a ''complete physical theory'' and
the conclusions of the standard quantum mechanics do not
irreconcilably contradict each other as is commonly believed. In
the algebraic approach, we formulate axioms that allow
constructing a renewed mathematical scheme of quantum mechanics.
This scheme involves the standard mathematical formalism of
quantum mechanics. Simultaneously, it contains a mathematical
object that adequately describes a single experiment. We give an
example of the application of the proposed scheme.
\end{abstract}

\section{Introduction}

 In the seminal work by Einstein, Podolsky, and Rosen~\cc{epr},
 the main principles were formulated that must be satisfied by a
complete physical theory in the authors' opinion:

 a. ''each element of the physical reality must have a counterpart in the
  physical theory'' and

   b. ''if we can confidently (i.e., with the probability
one) predict the value of some physical quantity without
perturbing the system in any way, then there exists an element of
the physical reality corresponding to this physical quantity.''

 The standard quantum mechanics (the theory traced back to
Bohr, Heisenberg, Dirac, and von Neumann) did not accept this
maxim. A single experiment has no adequate counterpart in the
mathematical formalism of the standard quantum mechanics.
Moreover, the belief was firmly established that such a
counterpart cannot exist.

 In this work, we attempt to formulate
the basic points of a mathematical scheme of quantum mechanics
satisfying two seemingly incompatible requirements. First, the
standard mathematical formalism of quantum mechanics can be
reproduced within this scheme. Second, this scheme involves a
mathematical object that adequately describes a single experiment.

Most of these points are formulated in~\cc{slav}. We clarify and
extend them here. In particular, we essentially clarify the role
of the measuring device. In this work, in variation
with~\cc{slav}, greater attention is paid to mathematical aspects
of the theory. A more detailed phenomenological justification of
the axioms and the relations of the proposed approach to other
versions of the theory can be found in~\cc{slav}.

\section {Observables and physical states}

 The proposal is to construct the
mathematical formalism of quantum mechanics based on the algebraic
approach to quantum theory~\cc{emch}, where observable quantities
correspond to elements of some algebra. The main points of this
algebraic approach can be formulated more simply if in addition to
the directly observable quantities, their complex combinations are
also included in the consideration. These combinations are called
dynamic quantities in what follows. We accept the following
statement as the {\it first postulate}.

{\it Elements of an involutive, associative, and (in general)
noncommutative algebra~$\AAA$ correspond to dynamic variables such
that the following conditions are satisfied:

 a. for each element $\R\in\AAA$, there exists a Hermitian element
   $\A$ $(\A^*=\A)$ such that $\R^*\R=\A^2$, and

 b. if $\R^*\R=0$, then $\R=0$}.

  We assume that
the algebra has a unit element~$\I$ and that Hermitian elements of
$\AAA$ correspond to observable quantities. We let~$\AAA_+$ denote
the set of these elements.

 We now formulate {\it the second postulate}.

{\it Mutually commuting elements of the set~$\AAA_+$ correspond to
compatible (simultaneously measurable) observables}.

 A specific
feature of compatible observables is then that they allow a system
of measuring devices whereby these observables can be repeatedly
measured in an arbitrary sequence. The results of a repeated
measurement of observables are then unchanged. We say that the
corresponding measurements are reproducible.

 We let $\QQQ \quad (\QQ\equiv \{\Q\}\in\AAA_+)$
denote the maximal real commutative subalgebra of the
algebra~\AAA. This is the subalgebra of compatible observables.
The subscript $\xi$ $(\xi \in \Xi)$ distinguishes different such
subalgebras. If $\AAA$ is commutative (the algebra of classical
dynamic variables), then the set $\Xi$ consists of one element. If
$\AAA$ is noncommutative (the algebra of quantum dynamic
variables), then the set $\Xi$ has the power of the continuum.

Hermitian elements of~\AAA{} are a latent form of observable
quantities. The explicit form is a certain number, which
transpires in an {\it individual} observation. We assume that
there exists some physical reality that determines the result of
such an individual observation. We call this physical reality the
physical state of the quantum object.

 In what follows, we need the following
definition~\cc{rud}. Let \BB{} be a real (complex) commutative
algebra and $\tvp$ be a linear functional on~\BB. If
$$\tvp(\B_1\B_2)=\tvp(\B_1)\tvp(\B_2) \mbox { for all }\B_1\in\BB
\mbox { and } \B_2\in\BB,$$
 then the functional $\tvp$ is called the
real (complex) homomorphism on the algebra \BB.

 We now formulate the {\it third postulate}.

 {\it The physical state of a quantum object
involved in an individual observation is described by a functional
$\vp(\A)$  (in general, multivalued), with $(\A\in\AAA_+)$, whose
restriction $\vpx$ to each subalgebra~$\QQQ$ is single-valued and
is a real homomorphism ($\vpx=A$ is a real number).}

A functional $\vp$ is multivalued because the result of an
observation can depend not only on the quantum object under
observation but also on properties of the device used for the
observation. A typical measuring device consists of an analyzer
and a detector. The analyzer is a device with one input and
several output channels. As an example, we consider the device
measuring an observable $\A$. For simplicity, we assume that the
spectrum of this observable is discrete. Each output channel of
the analyzer must then correspond to a certain point of the
spectrum. The detector registers the output channel through which
the quantum object leaves the analyzer. The corresponding point of
the spectrum is taken to be the value of the observable $\A$
registered by the measuring device.

In general, the value not of one observable $\A_i$ but of an
entire set of compatible observables can be registered in one
experiment. All these observables must belong to a single
subalgebra $\QQQ$. Each output channel of the analyzer must then
correspond to a set of points of the spectra of the observables
$\A_i$, one point for each independent observable. Obviously, the
analyzer must be constructed appropriately. The analyzer
constructed in this way (and the entire measuring device) is said
to be compatible with the subalgebra $\QQQ$.

The set of observables $\A_i$ need not necessarily be registered
by one device. A group of measuring devices (one complicated
device) can be used for this purpose. In this case, the entire
group (each of its elements) must be compatible with the
subalgebra $\QQQ$. We assume that the restriction $\vpx$ of $\vp$
corresponding to a certain physical state describes the value of
the observable $\A$ that is registered in this physical state by
the measuring device compatible with the subalgebra $ \QQQ $.

We assume that the compatibility of a device with a subalgebra is
determined by classical characteristics (construction, spatial
position, etc.) of the device. If the device is compatible with a
subalgebra $\QQQ$, we say that it belongs to the type $ \QQQ $.

One observable can belong to two (and more) different subalgebras,
$\A\in \QQQ\cap\QQ_{\xi'} $. If the functional $\vp$ is
multivalued at a point $\A$, it can occur that
$\vpx\ne\vp_{\xi'}(\A)$. Physically, this means that we can
register different values of the same observable in the same
physical state using different measuring devices. Therefore, the
functional $\vp$ does not describe the value of the observable
$\A$ in a certain physical state. It describes the response of the
measuring device of a certain type to the observable $\A$.
Accordingly, the physical reality is not the value of the
observable $\A$ in a given physical state but the response of the
measuring device to this state.

If the functional $\vp$ is single-valued at a point $\A$, we say
that the corresponding physical state $\vp$ is stable on the
observable $\A$. In this case, we can say that the observable $\A$
has a definite value in the physical state $\vp$ and this value is
the physical reality.

The functionals involved in the third postulate can be shown to
have the following properties~\cc{rud}:
 \bea{3} &/1/& \vx(0)=0; \nn{} &/2/& \vx(\I)=1; \nn{}
&/3/& \vx(\A^2)\ge 0; \nn{} &/4/& \mbox{if } \lll=\vx(\A), \mbox{
then } \lll\in\sss(\A); \nn{} &/5/& \mbox{if } \lll\in\sss(\A),
\mbox{ then } \lll=\vx(\A) \mbox{ for some } \vx(\A). \eea
 Here, $\sss(\A)$ is the spectrum of the element $\A$ in the algebra \AAA.
 The corresponding properties of individual measurements are
postulated in the standard quantum mechanics but are a consequence
of the third postulate here.

The multivaluedness of a functional $\vp$ allows introducing it
consistently. This can be verified by direct construction.
Evidently, it suffices to construct the restriction $\vx$ of $\vp$
to each subalgebra $\QQQ$.

We describe several ways of constructing the functional $\vp$. The
first is as follows. In each subalgebra $\QQQ$, we arbitrarily
choose a system $G(\QQQ)$ of independent generators. We next
require $\vx$ to be a certain mapping of $G(\QQQ)$ to a real
number set $S_{\xi}$ (allowable points of the spectra for the
corresponding elements of the set $G(\QQQ)$). On the other
elements of $\QQQ$, the functional $\vx$ is constructed by
linearity and multiplicativity.

It is clear that this procedure is always possible if each
functional $\vx$ is constructed independently of the others. On
the other hand, the functional $\vp$ resulting from this
construction is highly ambiguous.

We can attempt constructing a single-valued functional $\vp$. For
this, we choose some subalgebra $\QQ_1$ (of type \QQ ) and let
$G(\QQ_1)$ be a set of generators of $\QQ_1$. We define the
restriction $\vp_1$ of  $\vp$ to the subalgebra $\QQ_1$ by
requiring $\vp_1$ to be some mapping of $G(\QQ_1)$ to a real
number set $S_1$ (points of the spectra for the corresponding
elements of $\QQ_1$).

We next choose another subalgebra $\QQ_2$. With $\QQ_1\cap
\QQ_2\equiv \QQ_{12}\ne \emptyset$, we first construct a set of
generators  $ G_{12}$ of $\QQ_{12}$, and then supplement it with
the set  $G_{21}$ to the complete set of generators of $\QQ_2$.
The restriction $\vp_2$ is constructed as follows. If $\A\in
G_{12}$, then $\vp_2(\A)=\vp_1(\A)$. If $\A\in G_{21}$, then the
functional $\vp_2$ is defined such that it is a mapping of $
G_{21}$ to some allowable set of points in the spectra of the
corresponding elements of the algebra  $\QQ_2$.

The same scheme is used to construct the restrictions of the
functional $\vp$ to other subalgebras $\QQQ$. But this scheme can
become inconsistent at a certain stage because the subalgebra
$\QQQ$ can have nonempty intersections with other subalgebras
$\QQ_{\xi1}, \QQ_{\xi2}, \dots , \QQ_{\xi n}$ for which the
restrictions of $\vp$ are already fixed. The corresponding
mappings can be nonallowable for elements of the subalgebra
$\QQQ$.

In this case, we proceed as follows. Let $k$ be the maximum number
($1\le k \le n $) such that the definitions $\vpx=\vp_{\xi i}(\A)$
are allowable for all $ i $ ($1\le i \le k$). With these
definitions, we define the functional $\vpx$ on the subalgebra
$\QQ^{k}=[\QQ_{\xi1}\cup\dots \cup \QQ_{\xi k}]\cap\QQQ$. We next
take some set $G^{(k)}$ of generators of the subalgebra $\QQ^k$
and supplement it with the set $G_k$ to a complete set of
generators of $\QQQ$. On the generators in $G_k$, the functional
$\vpx$ is defined such that it realizes the mapping of $G_k$ to
the set of allowable points of the spectra of the corresponding
elements of $\QQQ$. At this stage, the functional $\vp$ can become
multivalued, and a single-valued functional therefore cannot be
constructed. But the arising ambiguity is minimal in a certain
sense. It is therefore impossible to avoid ambiguity in the
functional $\vp$ in the general case. Kochen and Specker~\cc{ksp}
have given a specific example of this case.

But it is always possible to construct a functional $\vp$ that is
single-valued on all observables belonging to any preset
subalgebra $\QQQ$. For this, it suffices to assign the subalgebra
$\QQQ$ number 1 (set $\xi=1$) and define the restriction $\vp_1$
of $\vp$ to $\QQ_1$ as described above. We must next exhaust all
subalgebras $\QQ_i$ (of type $\QQQ$) that have nonempty
intersections with $\QQ_1$. To construct the restriction $\vp_i$
of $\vp$ to each $\QQ_i$, it suffices to use the recipe used in
the previous version in constructing the restriction $\vp_2$. By
construction, such a functional $\vp$ is single-valued on all
elements belonging to $\QQ_1$. Different subalgebras $\QQ_i$ can
have common elements that do not belong to $\QQ_1$. On these
elements, the functional $\vp$ can be multivalued.

 We accept the following statement as the {\it fourth postulate}.

{\it The equality $ \vx(\A_1)=\vx(\A_2)$ is satisfied for all
$\vx$ if and only if $\A_1=\A_2$.}

 In other words, the functional $\vp$ separates arbitrary two different
 observables. The equality $\A_1=\A_2$ in particular denotes
 that both elements $\A_1$ and $\A_2$ simultaneously belong
 (or do not belong) to a domain of the functional $\vx$.

 \section {The quantum ensemble}

  The functional $\vp$ maps the set $\QQ_{\xi}=\{\Q\}_{\xi}$ into a real
number set,
 $$
  \{\Q\}_{\xi}\stackrel{\vp}{\longrightarrow}\{Q=\vp(\Q)\}_{\xi}.
 $$
  For different functionals $\vp_i$ and $\vp_j$, the sets
   $\{\vp_i(\Q)\}_{\xi}$ and $\{\vp_j(\Q)\}_{\xi}$  can be different or
can coincide. If $\vp_i(\Q)=\vp_j(\Q)=Q $ for all $\Q \in \{\Q\}$,
then the functionals $\vp_i$ and $\vp_j$ are said to be
$\{Q\}$-equivalent. We let $\{\vp\}_Q$ be the set of all physical
states to which there correspond $\{Q\}$-equivalent functionals
that are stable on the observables in the subalgebra
$\{\Q\}_{\xi}$. The set of the corresponding physical states is
said to be a (pure) quantum state and is denoted by~$\Psi_Q$. The
set of physical systems that are in these physical states is said
to be the quantum~$\Psi_Q$-ensemble.

Strictly speaking, the above definition of the quantum state is
only valid for a physical system that does not contain identical
particles. Describing identical particles requires some
generalization of the definition of the quantum state~\cc{slav}.

We consider a quantum~$\Psi_Q$-ensemble as a parent population (in
the probability theory sense) and each experiment to measure an
observable $\A$ as a trial. As the event $\aA$, we consider the
experiment where the measured value of the observable $\A$ is not
greater than $\aA$, i.e., $\vp(\A)=A\le\aA$. This event is not
unconditional. By the second postulate, one trial cannot be an
event for two noncommuting observables. The probability of the
event $\aA$ is determined by the structure of the quantum ensemble
and by this condition. Let this probability be equal to $\pP$.

 We let $\{\vp\}^{\A}_Q$ (with ($\{\vp\}^{\A}_Q\subset\{\vp\}_Q$)
 denote the set of physical states involved in a denumerable sampling
from mutually independent random trials of measuring the
observable $\A$. We note that if $\B$ is an observable not
commuting with $\A$, then the probability that the sets
$\{\vp\}^{\A}_Q$ and $\{\vp\}^{\B}_Q$  intersect is equal to zero.
Indeed, on the one hand, the observables $\A$ and $\B$ cannot be
measured in one trial. On the other hand, the set $\{\vp\}_Q$ has
the power of the continuum. Therefore, the probability that the
same state from $\{\vp\}_Q$ is repeated in two random denumerable
samplings is equal to zero. Therefore, the additional condition is
automatically satisfied with the probability one for the described
samplings.

 By definition, the probability of the occurrence of an
event $\aA$ in each of these trials is $\pP$. It determines the
probabilistic measure $\mu(\vp)$  $(\vp(\A)\le \aA)$ on any such
sampling. The measure $\mu(\vp)$ in turn determines a distribution
of the values $A_i=\vp_i(\A)$ of the observable $\A$ and the
mathematical expectation $<A>$ in this sampling,
$$<A>=\int_{\{\vp\}^{\A}_Q}d\mu(\vp)\,\vp(\A).$$

 For $1\le i \le n$, let the functionals $\vp_i \in\{\vp\}^{\A}_Q$, then by
Khinchin's theorem (the law of large numbers; see,
e.g.,~\cc{nev}), the random quantity $\bar A_n=(A_1+\dots+A_n)/n$
converges to $<A>$ in probability as $n\to\infty$. Therefore,
\beq{8}
 \mbox{P-}\lim_{n\to\infty}\frac{1}{n}
\Big(\vp_1(\A)+\dots+\vp_n(\A)\Big) = <A>\equiv\Psi_Q(\A).
 \eeq

Formula \rr{8} defines a functional (the quantum average) on the
set $\AAA_+$. The totality of all quantum experiments
unambiguously indicates that we must accept the following {\it
linearity postulate}.

{\it The functional $\Psi_Q(\;)$ is linear on the set $\AAA_+$.}

This implies that $$ \Psi_Q(\A+\B)=\Psi_Q(\A)+\Psi_Q(\B)$$ also in
the case where $[\A,\B]\ne0$.

Quantum experiments support one more postulate. To determine the
probability $\pP$ experimentally, we must conduct random tests.
But these tests are often accompanied in practice by some
condition because the quantum average of an observable is
experimentally found using not a random set of devices capable of
measuring this observable but a certain type of such devices.
Devices of different types can be used in different series of
experiments. Experience shows that the probability $\pP$ is the
same in all these cases. The {\it representativity postulate} is
therefore true.

{\it The probability $\pP$ of detecting an event $\aA$ for a
system in any quantum state $\Psi_Q$ is independent of the type of
the measuring device used for that purpose.}

 Obviously, an ideal measuring device is understood here. Any realistic
 device introduces a systematic error.

We now discuss how the representativity postulate requirement can
be realized in the proposed approach. Let a functional $\vp^{\mu}$
be multivalued on an observable $\A$. This is related to the fact
that the result of the observation of a certain quantum object can
depend not only on the internal properties of this object but also
on the type of the device intended for investigating the
observable $\A$. We label different types of devices in some fixed
manner. Let $\{D_1(\A),D_2(\A),\dots,D_n(\A)\}$ be the set of
different types of devices. We label the values of a multivalued
functional $\vp^{\mu}$  on the elements $\A$ in accordance with
the labeling of the measuring devices,
 \beq{81}
  \vp^{\mu}(\A)=\{A_{\mu1},\dots,A_{\mu n}\},
  \eeq
where $A_{\mu i}$ is the indication of the device $D_i(\A)$ in the
physical state $\vp^{\mu}$. Each specific device $D_i(\A)$ has
entirely definite physical characteristics and is therefore
compatible with a certain subalgebra $\QQ_i$ ($\A\in\QQ_i$). In
this sense, it ''knows'' which value of the functional to choose.
For a specific physical state, some (or all) values $A_{\mu i}$
can coincide.

 We next consider another functional $\vp^{\nu}$, whose
values on the observable $\A$ are
 \beq{82}
 \vp^{\nu}=\{A_{\nu1},\dots,A_{\nu n}\},
 \eeq
where $\nu1,\dots,\nu n$ is a certain permutation of
$\mu1,\dots,\mu n$ and $\quad A_{\nu i}$ denotes the indication of
the device $D_i(\A)$ in the physical state $\vp^{\nu}$. Let the
functionals $\vp^{\mu}$  and $\vp^{\nu}$ coincide on the other
observables.

 If a physical state $\vp^{\mu}$ belongs to the quantum state
$\Psi_Q=\{\vp\}_Q$, then the physical state $\vp^{\nu}$ also
belongs to $\Psi_Q$. Indeed, if $\A\in\{\Q\}$, then the
functionals $\vp^{\mu}$  and $\vp^{\nu}$ are single-valued on $\A$
and therefore coincide. If $\A\notin\{\Q\}$, then the functionals
$\vp^{\mu}$  and $\vp^{\nu}$ coincide on $\Q\in\{\Q\}$ by
construction. This argument holds if we consider the functionals
obtained by any other permutation of the indices $\mu1,\dots,\mu
n$.

 We now verify that for the representativity postulate to be
satisfied, it suffices to require the relative probabilities of
hitting the states $\vp^{\mu}$ and $\vp^{\nu}$ to be the same.
Indeed, let the device $D_i(\A)$ be used to find the mean values
of the observable $\A$ in a quantum state $\Psi_Q$ experimentally
and let $\Psi_Q^i(\A)$ denote the result of this experiment.
Theoretically, the quantity $\Psi_Q^i(\A)$ is constructed as
follows. We take some physical state $\vp^{\mu}$. This state
contributes $A_{\mu i}$ to $\Psi_Q^i(\A)$ with the weight
$w_{\mu}$ (the probability of hitting the state $\vp^{\mu}$). We
next take the state $\vp^{\nu}$. It contributes $\A_{\nu i}$ with
the weight $w_{\nu}=w_{\mu}$. In the same way, we must take all
the states obtained by other permutations of the arguments in
\rr{81} into account. Following the same scheme, we must then take
all the physical states that are not related to $\vp^{\mu}$ by a
permutation of the arguments in Eq.~\rr{81} into account.

 Now let the device $D_j(\A)$ be used to find the mean value. The result of
this experiment is denoted by $\Psi_Q^j(\A)$. To calculate
$\Psi_Q^j(\A)$ theoretically, we use the above scheme. We again
start with the states $\vp^{\mu}$, $\vp^{\nu}\dots$. We then
obtain a set of values $A_{\mu j}, A_{\nu j}\dots$ that contribute
to $\Psi_Q^j(\A)$ with the weight $w_{\mu}$. But the sets ($A_{\mu
i}, A_{\nu i}\dots$) and ($A_{\mu j}, A_{\nu j}\dots$) differ by
only a permutation of elements. Therefore, the total contributions
of the physical states $\vp^{\mu}$, $\vp^{\nu}\dots$ to
$\Psi_Q^i(\A)$ and $\Psi_Q^j(\A)$) are the same. Similarly, we
consider the contributions of the physical states that are not
related to $\vp^{\mu}$ a permutation of arguments. It follows that
$\Psi_Q^i(\A)=\Psi_Q^j(\A)$.

For the representativity postulate to be satisfied, it now
suffices to verify that for each physical state $\vp^{\mu}$, there
exists the corresponding state of the type $\vp^{\nu}$. Obviously,
this is indeed so if one additional point is introduced into the
constructive scheme of building the set of physical states
described above. Along with each multivaled functional $\vp^{\mu}$
constructed in accordance with the proposed scheme, all the
functionals whose values are obtained by all possible permutations
of the arguments in Eq.~\rr{82} must be introduced into the set of
functionals.

 Therefore, although the devices $D_i(\A)$ and $D_j(\A)$ can give
different results for the same physical state in individual
observations, they give the same result for mean values of the
observable $\A$. The proposed model thus gives the same results
for determining quantum mean values of observables as the standard
quantum mechanics.

Any element $\R$ of the algebra \AAA{} is uniquely represented as
$\R=\A+i\B$, where $\A,\B\in\AAA_+$. Therefore, the functional
$\Psi_Q$ can be uniquely extended to a linear functional on \AAA:
$\Psi_Q(\R)=\Psi_Q(\A)+i\Psi_Q(\B)$.

  Let  $\R^*\R=\A^2\in\{\Q\}$ ($\A\in\AAA_+$) If $\vp\in\{\vp\}_Q$,
  then $\vp(\R^*\R)=A^2$. Therefore,
 $$
\lt.\Psi_Q(\R^*\R)=A^2=\vp(\A^2)\rt|_{\vp\in\{\vp\}_Q}. $$

In accordance with inequality (1 /3/), we have $A^2\ge0$. In
addition, it follows from condition (1 /1/) and the fourth
postulate that $\sup\vp(\R^*\R)>0,\mbox{ if }\R\ne0$.

We define the norm of an element $\R$ by
 \beq{9}
 \|\R\|^2=\sup_{\vp}\vp(\R^*\R) =\sup_Q \Psi_Q(\R^*\R).
 \eeq
Because $\Psi_Q$ is a positive linear functional, all the axioms
of a norm are indeed satisfied for $\|\R\|$ (see~\cc{slav,emch}).
Because  $\vp([\A^2]^2)=[\vp(\A^2)]^2$, we have
$\|\R^*\R\|=\|\R\|^2$, and~\AAA{} is therefore a $C^*$-algebra.
Thus, a necessary condition for the consistency of the linearity
postulate is the following requirement: the algebra \AAA{} can be
endowed with the structure of a $C^*$-algebra.

This requirement can be formulated in purely algebraic terms
because the relation~\cc{dix}
 \beq{10}
 \rho(\R)=\|\R\|=\rho^{1/2}(\R^*\R).
 \eeq
is valid for a $C^*$-algebra. Here, $\rho(\R)$ is the spectral
radius of the element $\R$, $\rho(\R)=\sup_{\lambda}|\lambda_R|$,
where $\lambda_R\in\sigma(\R)$.

Using Eq.(1 /5/), we can rewrite relation~\rr{10} as
 $$
\|\R\|=[\sup_{\vp}\vp(\R^*\R)]^{1/2}. $$
 which agrees with
Eq.~\rr{9}. Therefore, the spectral radius (a purely algebraic
notion) of each element of the algebra \AAA{} must satisfy the
norm axioms and the condition $\rho^2(\R)=\rho(\R^*\R)$.

In view of the above, it is useful to give a new formulation of
the {\it first postulate}.

{\it Elements of the algebra \AAA{} that has the structure of a
$C^*$-algebra correspond to dynamic quantities.}

We did not accept this formulation of the first postulate
initially because it follows directly from the experiment that the
observables have algebraic properties and the quantum mean values
have the linearity property. But the mathematical relations
involved in the definition of a $C^*$-algebra are not directly
related to the experiment.

Postulating that the observables belong to a normed algebra, we
must consider the observables to be bounded. In any experiment, we
always deal with bounded values of observables. Normalizability of
the algebra is therefore not a restriction from the experimental
standpoint. But many unbounded operators, which are obviously not
elements of a normed algebra, occur in the quantum theory. In the
algebraic approach, ''unbounded observables'' are conventionally
considered as elements adjoint to the algebra of bounded dynamic
quantities; in other words, unbounded observables are assumed to
admit a spectral representation where the spectral projectors are
elements of a normed algebra. The ensuing problems are common to
the algebraic approach to quantum theory in general. We do not
consider them here.

\section{Time evolution and the ergodicity condition}

In the  standard quantum mechanics, the time evolution is
determined by the Heisenberg equation
 \beq{12}
\frac{d\,\A(t)}{d\,t}=i\Big[\HH,\A(t)\Big], \qquad \A(0)=\A,
 \eeq
 where $\A(t)$ and the Hamiltonian $\HH$ are operators in some
Hilbert space. But for~\rr{12} to preserve its physical meaning,
it suffices to consider $\A(t)$ and $\HH$ as elements of some
algebra (in particular, of \AAA{}) or elements adjoint to the
algebra.

 In our case, the evolution equation can be rewritten in
terms of physical states. We therefore accept the {\it fifth
postulate}.

 {\it A physical state of a quantum system evolves in time
as
 \beq{13}
 \vp(\A) \to \vp_t(\A)\equiv\vp(\A(t)),
 \eeq
where $\A(t)$ is defined by Eq.~\rr{12}.}

Equation~\rr{13} describes time evolution of a physical state
entirely unambiguously. It is a different story, though, that an
observation allows determining the initial value $\vp(\A)$ of a
functional only up to its belonging to a certain quantum state
$\{\vp\}_Q$~\cc{slav}. Most of our predictions regarding the time
evolution of a quantum object are therefore probabilistic. In
addition, Eqs.~\rr{12} and~\rr{13} are valid only for systems that
are not exposed to first-class actions (in von Neumann's
terminology~\cc{von}), i.e., do not interact with a classical
measuring device.

We now return to the linearity postulate. From the experimental
standpoint, this postulate is well justified. But it is not quite
clear whether it can be realized within the mathematical scheme
considered here. It turns out that this postulate can be related
to the time evolution of the quantum system. For this, we must
impose restrictions on the Hamiltonian $\HH$.

We now accept the {\it sixth postulate}.

{\it The Hamiltonian $\HH$ is a Hermitian spectral (possibly,
adjoint) element of the algebra \AAA. The spectrum of $\HH$
contains at least one discrete nondegenerate value $E_0$. }

This implies that the Hamiltonian $\HH$ has an integral
representation of the form
 \beq{014}
\HH=\int \p(dE)\;E, \eeq where $\p(dE)$ are orthogonal projectors.
Hereinafter integrations (and also limits) on algebra~\AAA{} are
understood in sense of the weak topology of $C^*$-algebra.

Somewhat conventionally, we can represent $\p(dE)$ as
 \beq{14}
 \p(dE)=\p_p(dE)+\p_c(dE)=\sum_n \p_n\,\delta(E-E_n)\,dE+
 \p_c(dE).
 \eeq
Here $\p_p(dE)$ and $\p_c(dE)$ concern to point and continuous
spectrums, correspondingly. Besides, $\p_n\p_m=\p_m\p_n=0$ for
$m\ne n$, $\p_n\p_c(dE)=\p_c(dE)\p_n=0$. The sum over $n$
in~\rr{14} must necessarily involve at least one term ($n=0$) with
a nondegenerate value $E_0$.

In addition to this last restriction, other requirements are
always assumed in considering any quantum mechanics model.
Requiring a discrete point in the spectrum does not seem too
restrictive either. For example, a one-particle quantum system can
have a purely continuous energy spectrum. But it can be considered
as a one-particle state of an extended system that can also be in
the vacuum state in addition to the one-particle state. The energy
spectrum of the extended system already has a discrete
nondegenerate point in the spectrum.

The quantity $E_0$ need not necessarily be the lower bound of the
spectrum. By the nondegeneracy of $E_0$, we assume that the
projector $\p_0$ in decomposition~\rr{14} is one-dimensional. A
projector $\p$ is said to be one-dimensional if it cannot be
represented as
$$\p=\sum_{\al}\p_{\al}, \quad \p_{\al}\ne\p, \quad
 \p\p_{\al}=\p_{\al}\p=\p_{\al}.$$

Let us remark that, if two elements $\A_1$ and $A_2$ of
algebra~\AAA{} have identical spectral representation of
type~\rr{014}, then they obey the fourth postulate. Therefore,
such elements coincide.

{\bf Statement 1.}

{\it If $\A\in\AAA_+$, then $\A_0=\p_0\A\p_0$ has the form
$\A_0=\p_0\,\po(\A)$, where $\po(\A)$ is some functional.}

{\bf Proof.} Because $[\A_0,\p_0]=0$, it follows that $\A_0$ and
$\p_0$ have the common spectral decomposition of unity. Because
the projector $\p_0$ is one-dimensional, the spectral
decomposition of $\A_0$ must have the form
$\A_0=\p_0\,\po(\A)+\A'_0$, where $\A'_0$ is orthogonal to $\p_0$.
Therefore, $\A_0=\p_0\A_0=\p_0\p_0\po(\A)+
 \p_0\A'_0=\p_0\po(\A)$. The statement is proved.

 {\bf Statement 2.}

{\it The functional $\po(\A)$ is linear.}

{\bf Proof.} Indeed,
 $$ \p_0\po(\A+\B)=\p_0(\A+\B)\p_0=\p_0\,\po(\A)+\p_0\po(\B).$$
  Because $\p_0\ne 0$, it follows that
  $\po(\A+\B)=\po(\A)+\po(\B)$, which was to be proved.

A physical state $\vp_{0\al}$ is said to be ground if
$\vp_{0\al}(\p_0)=1$. By linearity, the functional $\po(\A)$ is
uniquely extended to the algebra \AAA,
$\po(\A+i\,\B)=\po(\A)+i\po(\B)$, where $\A,\B\in\AAA_+$.

{\bf Statement 3.} {\it The functional $\po(\A)$ is positive.}

{\bf Proof.} In accordance with property (1 /3/), there is the
inequality $\vp_{0\al}(\p_0\R^*\R\p_0)\geq 0$. On the other hand,
$$\vp_{0\al}(\p_0\R^*\R\p_0)=\vp_{0\al}(\p_0\po(\R^*\R))
 =\po(\R^*\R)$$. The statement is proved.

 {\bf Statement 4.}
{\it The functional $\po$ satisfies the normalization condition
$\po(\I)=1$.}

{\it Proof.} Indeed,
$$1=\vp_{0\al}(\p_0\I\p_0)=\vp_{0\al}(\p_0\po(\I))=\po(\I),$$
which was to be proved.

To find the physical meaning of the functional $\po$, we consider
an element $\Aa$ in the algebra \AAA{} that corresponds to an
observable $\A$ averaged in time,
 \beq{15}
 \Aa=\lim_{L\to\infty}\frac{1}{2L}\int^L_{-L}dt\,\A(t)=
    \lim_{L\to\infty}\frac{1}{2L}\int^L_{-L}dt\,\exp[-i\HH t]
    \,\A\,\exp[i\HH t].
 \eeq
The average is understood with respect to the weak topology of
$C^*$-algebra. Substituting the spectral decomposition of $\HH$ in
\rr{15}, we  obtain
 $$
\Aa=\lim_{L\to\infty}\frac{1}{2L}\int^L_{-L}dt\,\left[\sum_n
\p_n\A\p_n +\int\,\p_c(dE)\A\p_c(dE')\,\exp[i(E'-E)]+
\B(t)\right], $$
 where
\begin{eqnarray*}
  \B(t)&=& \sum_n\left[\p_n\,\A\int\,\p_c(dE)\,\exp[i(E_n-E)t]+
  \p_c(dE)\,\A\,\p_n\, \exp[-i(E_n-E)t]\right]\\
  &+& \sum_{n,m}^{n\ne m}\p_n\,\A\,\p_m \exp[i(E_n-E_m)t]
\end{eqnarray*}

 It can be easily shown that the occurrence
of time exponentials in this expression implies the relation
$$\lim_{L\to\infty}\frac{1}{2L}\int^L_{-L}dt\,\B(t)=0,
 $$
  and therefore
 $$ \Aa=\sum_n\p_n\,\A\,\p_n+\D, $$
 where
$$ \D=\lim_{L\to\infty}\frac{1}{2L}\int^L_{-L}dt\,
\int\,\p_c(dE)\A\p_c(dE')\,\exp[i(E'-E)].$$ In a right-hand side
of this equality the integrand $\A\exp[i(E'-E)]$ is majorized by
magnitude $\|\A\|$. Therefore, the integrals and the limit exist.
Because $\p_c(dE)=\p_c\p_c(dE)=\p_c(dE)\p_c$, where
$\p_c=\int\,\p_c(dE)$, we have $\D=\p_c\D\p_c$. Therefore,
 $$ \Aa=\sum_n\p_n\,\A\,\p_n+\p_c\D\p_c. $$

We now consider the value of the observable $\Aa$ in the physical
ground state $\vp_{0\al}$,
 \beq{19}
 \vp_{0\al}\left(\sum_n\p_n\,\A\,\p_n+\p_c\D\p_c\right)=
 \vp_{0\al}(\p_0\,\A\,\p_0)+ \vp_{0\al}(\F),
 \eeq
 where $\F=\sum_{n\ne 0}\p_n\,\A\,\p_n+\p_c\D\p_c$.
We here use the linearity of the functional $\vp_{0\al}$ on
mutually commuting elements $\p_n\,\A\,\p_n $,
 $\p_m\,\A\,\p_m $ and $\p_c\D\p_c$.
Because $\p_n\p_0=\p_c\p_0=0$ for $n\ne 0$, the right-hand side of
\rr{19} can be rewritten as
 $$
\po(\A)+\vp_{0\al}\left((\I-\p_0)\F(\I-\p_0)\right)
=\po(\A)+\vp_{0\al}(\I-\p_0)\vp_{0\al}\left((\I-\p_0)\F
   (\I-\p_0)\right)=\po(\A).
   $$
We finally have  $$ \vp_{0\al}(\Aa)=\po(\A). $$ The value of the
observable $\Aa$ is the same in all physical ground states.

 The functional $\po$ has all the properties that must
be possessed by a functional determining quantum mean values. It
is linear, is positive, and is equal to unity on the unit element.
In addition, it is continuous as a linear functional on the
$C^*$-algebra. Instead of the linearity axiom, we can therefore
accept the {\it seventh postulate} (the ergodicity axiom).

{\it The mean value of an observable $\A$ in the ground state of a
quantum ensemble is equal to the value of observable $\Aa$
(time-averaged value of the observable $\A$) in any physical
ground state.}

 To construct the standard mathematical formalism of
quantum mechanics, we can now use the canonical construction of
Gelfand-Naimark-Segal (GNS)(see, e.g.,~\cc{emch}).

We consider two elements $\R,\,\s\in\AAA$ equivalent if the
condition $\po\left(\K^*(\R-\s)\right)=0$ is valid for any
$\K\in\AAA$. We let $\Phi(\R)$ denote the equivalence class of the
element $\R$ and consider the set $\AAA(\po)$ of all equivalence
classes in \AAA. We make $\AAA(\po)$ a linear space setting
$a\Phi(\R)+b\Phi(\s)=\Phi(a\R+b\s)$. The scalar product in
$\AAA(\po)$ is defined as
$\left(\Phi(\R),\Phi(\s)\right)=\po(\R^*\s)$. This scalar product
generates the norm $\|\Phi(\R)\|^2=\po(\R^*\R)$ in  $\AAA(\po)$.
Completion with respect to this norm makes $\AAA(\po)$ a Hilbert
space. Each element $\s$ of the algebra \AAA{} is uniquely
assigned a linear operator $\Pi_{\Psi}(\s)$ acting in this space
as $\Pi_{\Psi}(\s)\Phi(\R)=\Phi(\s\R)$.

We note that using the ground state $\po$ is not a necessary
condition in the GNS construction. All our argument can be based
not on the projection operator $\p_0$ but on any other
one-dimensional projection operator $\p$. In this case, there is
also the equality
 \beq{21} \p\A\p=\p\Psi_p(\A), \eeq
where $\Psi_p(\A)$ is a functional with the linearity and
positivity properties; in addition, $\Psi_p(\I)=1$.

Instead of the physical ground states $\vp_{0\al}$, we can use the
basic physical states $\vp_{p\al}$. These states satisfy the
condition $\vp_{p\al}(\p)=1$. The set $\{\vp_{p\al}\}$ of physical
states constitutes the base quantum state $\{\vp\}_p$. Instead of
the ergodicity axiom, we can postulate that the quantum average in
the state $\{\vp\}_p$ is defined by the functional $\Psi_p$ in
Eq.~\rr{21}. In this case, we lose the relation of the quantum
average to the time average. As a compensation, the theory becomes
more flexible because the assumptions regarding the spectrum of
the Hamiltonian become redundant.

 \section{An example}
To illustrate the above, we consider a quantum system whose
observable quantities are described by Hermitian $2\times 2$
matrices. The elements $\HH$, $\p_0$, and $\A$ are given by
$$\HH=\lt[ \begin{array}{cc}
  E_0 & 0 \\
  0 & -E_0
\end{array}\rt],\qquad \p_0=\lt[ \begin{array}{cc}
  0 & 0 \\
  0 & 1
\end{array}\rt],\qquad \A=\lt[\begin{array}{cc}
  a & b \\
  c & d
\end{array}\rt]. $$
Obviously, $\p_0\,\po(\A)=\p_0\,\A\,\p_0=\p_0\,d$, i.e.,
 \beq{22}
 \po(\A)=d.
 \eeq
In addition, \beq{23}
 \Aa=\lim_{L\to\infty}\frac{1}{2L}\int^L_{-L}dt\,e^{-iE_0t\tau_3}
 \,\A\,e^{iE_0t\tau_3}=\lt[\begin{array}{cc}
   a & 0 \\
   0 & d
 \end{array}\rt],\eeq
 where $\tau_i$ are Pauli matrices.

All physical states can easily be constructed. We consider a
Hermitian matrix $\A$, i.e., with  $a^*=a$, $d^*=d$, and
 $c=b^*$. Any such matrix can be represented as
 \beq{24}
 \A=r_0\I+r\,\hat\tau(\bar n),
 \eeq
  where $\bar n$ is the unit three-dimensional vector, $\hat\tau(\bar
 n)=(\bar \tau\,\bar n)$. For Eq.\rr{24} to be valid, we must
  $$ r=\lt(\frac{(a-d)^2}{4}+b\,b^*\rt)^{1/2}, \quad
 r_0=\frac{a+d}{2}, $$ $$ n_1=\frac{b+b^*}{2r}, \quad
 n_2=\frac{b-b^*}{2ir}, \quad n_3=\frac{a-d}{2r}. $$
The commutator of the matrices $\hat\tau(\bar n),\hat\tau(\bar
n')$ is nonvanishing for $\bar n'\ne\pm\bar n$. Therefore, each
matrix $\hat\tau(\bar n)$ (up to a sign) is a generator of a real
maximal commutative subalgebra. Because $\hat\tau(\bar n)
\hat\tau(\bar n)=\I$, the spectrum of $\hat\tau(\bar n)$ consists
of two points $\pm1$.

Let $\{f(\bar n)\}$ be the set of all functions taking the values
$\pm1$ and such that $f(-\bar n)=-f(\bar n)$. A physical state is
described by a functional whose value coincides with one of the
points in the spectrum of the corresponding algebra element. For
each point of the spectrum, there exists an appropriate
functional. Therefore, to the set of physical states, there
corresponds a set of functionals defined by
$$ \vp\left(\hat
\tau(\bar{n})\right)=f(\bar{n}).$$
 Taking properties \rr{3} into
account (which must be possessed by each physical state), we
obtain
 \beq{26}
 \vp(\A)=r_0+r\,f(\bar{n}).
 \eeq
The ground state is any functional $\vp_{0\al}$ such that
$$f(n_1=0,n_2=0,n_3=1)=-1.$$
 Substituting the element $\Aa$
(\rr{23}) in \rr{26}, we obtain
 $$ \vp_{0\al}(\Aa)=\frac{a+d}{2}+\frac{a-d}{2}=d. $$
This agrees with \rr{22}.

In this model, we can do without multivalued functionals. If we
considered the algebra of matrices describing unit spin,
multivalued functionals would inevitably arise. This was noted (in
other terms) by Kochen and Specker \cc{ksp}.

\section{Conclusions}

In principle, a physical state can be considered as a special
hidden parameter. But because of the multivaluedness of the
functional $\vp$, the conditions of the Kochen and Specker no-go
theorem \cc{ksp} are not satisfied for this hidden parameter. It
was noted in \cc{slav} that the conditions of the von Neumann
no-go theorem \cc{von} are not satisfied for $\vp$. In addition,
it was also shown there that the conditions of Bell's theorem
\cc{bel} are not satisfied for the functional $\vp$. Therefore,
the arguments that are usually adduced by opponents of the use of
hidden parameters in quantum mechanics become inapplicable for the
physical state.

Regarding the multivaluedness of the functional $\vp$, the
physical process that is usually called a measurement should
rather be called an observation. The term ''observation'' better
expresses the fact that the indication of the device in use has
two causes: the physical state of the observed object and the type
of the device. The type of the device is then defined not only by
the observable that it must register but also by an ignored
parameter of the device. In the proposed approach, the type of the
measuring device plays the role of such an ignored parameter.
Different types of devices correspond to different maximal
commutative subalgebras to which a given observable belongs.
Unlike the values of a hidden parameter, the value of the ignored
parameter can in principle be established experimentally.

The result of an observation experiment (the physical reality) is
thus determined by two other physical realities: the physical
state of the observed object and the type of the observing device.
In an individual quantum experiment, in contrast to a classical
one, the second physical reality cannot be neglected in general.
But even in the quantum case, it is possible not to take the type
of the measuring device into account in an experiment determining
the mean values of an observable.

{\bf Acknowledgments.} The author is deeply grateful to
K.A.Sveshnikov and P.K.Silaev for a very valuable discussion.

\end{document}